\documentstyle[12pt,psfig] {article}
\begin{document}
\bibliographystyle{unsrt} 


\title{
\bf \Large Parameterisation of [$\sigma_{1/2}-\sigma_{3/2}$]  
for $Q^2\ge0$
and non-resonance contribution to the GDH sum rule
}

\author{N. Bianchi$^{\dagger}$ and E. Thomas\\
\small {\it INFN-Laboratori Nazionali di Frascati}\\
\small {\it CP 13, I-00044 Frascati, Italy}
}

\maketitle

\begin{abstract} 
A description of the virtual photon absorption cross section difference 
$\left[\sigma_{1/2}-\sigma_{3/2}\right]$ for the proton and neutron 
is obtained with a parameterisation based on a Regge type
approach.
The parametrisation is obtained from global fits to the cross section data 
derived from the spin asymmetries measured in deep inelastic scattering of longitudinally
polarised leptons from polarised $^1$H, $^3$He and $^2$H targets 
in the range 0.3 GeV$^2 < Q^2 <$ 70 GeV$^2$
and 4 GeV$^2 < W^2 <$ 300 GeV$^2$.
The fits give a reliable description of the data and provide predictions for the
photo-production through a smooth $Q^2$-transition. The contribution above 
the resonance region to the Gerasimov-Drell-Hearn sum rule for real and virtual photons has been evaluated.
For the real photons this contribution accounts for a large fraction of the discrepancy
between the sum rule expectations and the single pion photo-production analysis estimates.
\end{abstract}

\vspace{6.0cm}

\hspace{0.5cm}\footnotesize{$\dagger$ Corresponding author: Nicola.Bianchi@lnf.infn.it}

\vfill
\eject



\section{Introduction}

The study of deep-inelastic-scattering (DIS) of polarised leptons off polarised
nucleons provides information on the spin composition of the nucleon.
The experimental cross section asymmetry is generally written in terms of the target nucleon
polarised structure functions which have been successfully interpreted in terms of
parton spin distributions.
In this paper we analysed the available experimental data in terms of the virtual photon-nucleon
polarised cross sections which describe the multi-hadron polarised electroproduction.
This approach is complementary to the usual description
of the interaction of the probing virtual photon with the constituent partons.
It allows to extend the description of the process to the real photon case
in which the interactions can be unambiguously interpreted in terms of cross sections only.
The relevant quantity of our interest is the difference of the cross sections  
$\Delta\sigma = \left[\sigma_{1/2}-\sigma_{3/2}\right]$ in which $\sigma_{1/2}$
and $\sigma_{3/2}$ are the photon-nucleon absorption cross section of total helicity 1/2 and 
3/2, respectively.  

A fundamental sum rule for this quantity has been derived by 
Gerasimov, Drell and Hearn\cite{Geras} (GDH).
It relates the anomalous 
contribution $\kappa$ to the magnetic moment of the nucleon 
($\kappa_{\mathrm p}$=1.79, $\kappa_{\mathrm n}$=$-$1.91)
with the total absorption cross sections for circularly polarised real
photons on polarised nucleons.
It is given by:

\begin{equation}
\label{gdhorig}
I = \int_{\nu_0}^{\infty}
\Delta\sigma(\nu)\frac{d\nu}{\nu}=
-\frac{2\pi^2\alpha}{m^2}\kappa^2,
\end{equation}
where $\nu$ is the photon energy, $\nu_0$ is the pion photoproduction threshold
 and $m$ is the nucleon mass.
The theoretical predictions for the integral are --204 $\mu$b, --233 $\mu$b
and +29 $\mu$b for the proton, neutron and proton-neutron difference (p-n), respectively.
Experimentally this sum rule has never been tested directly because of the 
lack of suitable polarised targets and real photon beams. 
Several experiments are planned at different facilities to measure the spin-dependent
photo-production cross section in the nucleon resonance region and up to about 3 GeV\cite{bia}.
Multipole analysis of single pion photo-production amplitudes and estimates
of the double pion contribution from the nucleon resonance decays 
provided an indication of the low energy
contribution to the sum rule, which is expected to be the dominant one. 
These estimates were qualitatively consistent among
each others providing results ranging between --289  and --257 $\mu$b, --189 and --169
$\mu$b, --130 and --68 $\mu$b for the proton, neutron and proton-neutron
difference respectively[3-7]. All these findings 
strongly disagree with the GDH expectations 
and suggested a possible violation of the sum rule. 
In particular the multipole analysis of the isovector channel (i.e. the proton-neutron difference)
provides an opposite sign of the sum rule expectation.

The integral defined in Eq.~\ref{gdhorig} can be generalised to the absorption 
of virtual photons with energy $\nu$ and squared four-momentum $-Q^2$:

\begin{equation}
\label{gdhq2}
I(Q^2) = \int_{\nu_0}^{\infty}
\Delta\sigma(\nu, Q^2)\frac{d\nu}{\nu}.
\end{equation}
In the limit $Q^2 /\nu^2 \ll 1$ the above integral is
connected to the first momentum $\Gamma_1$ of the nucleon spin structure function
$g_1(x)$ over the Bjorken variable $x$ \cite{ANSELM}:

\begin{equation}
\label{gamma1}
I(Q^2) \approx I_1(Q^2) = \frac{16\pi^2\alpha}{Q^2} \Gamma_1 =
\frac{16\pi^2\alpha}{Q^2} \int_0^1 g_1(x)dx.  
\end{equation}
It is worth noting that in the transition region, from $Q^2$=0 to DIS regime, 
$I_1(Q^2)$ is only an approximation of $I(Q^2)$ \cite{sl}; therefore a 
determination of $[\sigma_{1/2}-\sigma_{3/2}]$ is needed for a precise 
evaluation of the $Q^2$ dependence of the generalised GDH integral, defined 
by Eq.~\ref{gdhq2}.
Several measurements are planned at TJNAF to measure the
resonance region contribution to the generalised  GDH integral $I(Q^2)$\cite{EX97}.

In this work we analysed the available polarised deep inelastic scattering data to evaluate
the contribution above the resonance region to the generalised  GDH integral $I(Q^2)$
and to provide an estimate for the real photon limit.
A Regge inspired parameterisation was used.
 
\section{Regge phenomenology}

The Regge theory provided a good description of the high-energy behaviour of the 
cross sections for a large number of processes\cite{DL}.
In this framework the cross section behaviours are given by $\sigma \sim
s^{\alpha^0-1}$ where $s$ is the centre of mass square energy  and 
$\alpha^0$ is the intercept of the leading Regge trajectory for the given process.
The intercept $\alpha^0 = J - \alpha' m_t^2$ is determined by the 
spin $J$ and mass $m_t$ of the exchanged particle and by the trajectory slope $\alpha' \simeq $ 0.8-0.9
GeV$^{-2}$.
In the case of the absorption of circularly polarised photons on polarised
nucleon it is useful to consider the isovector and the isoscalar terms of
the interaction\cite{Heim}.

The isovector contribution to $\Delta\sigma$
is described by the $a_1(1260)$ meson trajectory :
$\Delta\sigma_V \sim s^{\alpha^0_{a1}-1}$.
The axial vector meson $a_1$ is still now an elusive prey in most of 
experiments due to its large width and the presence of strong background.
Experiments of the last twenty years provided contradictory measurements of the $a_1$
meson mass and width over the wide ranges 1.05-1.28 GeV and 0.24-0.61 GeV, respectively\cite{pdg}.
The expected value of the $a_1$ meson intercept is about $-$0.3, but due to the
uncertainties in the knowledge of the mass and of the trajectory slope, a possible range between
0 and $-$0.5 is generally considered.

The isoscalar contribution to $\Delta\sigma$
is described by the $f_1(1285)$ meson trajectory 
$\Delta\sigma_I \sim s^{\alpha^0_{f1}-1}$. Being the $f_1$ meson mass
well known, its intercept is expected to be about $-0.4\pm0.1$.
In addition it has been shown that the exchange of two non-perturbative gluons
can provide an isoscalar contribution $\Delta\sigma_I \sim (1/s)ln(s/\mu^2)$ where the mass
parameter $\mu$ has the typical hadronic scale $\mu \sim 1$ GeV\cite{BL}.

In this work we consider the above described contributions 
for the case of photo-absorption of circularly polarised virtual photon on polarised nucleon
where the centre of mass square energy is given by $W^2=m^2+2m\nu-Q^2$.

\section{Data analysis and fitting procedure}

We have analysed the data from deep inelastic scattering of longitudinally
polarised leptons off polarised $^1$H, $^2$H and $^3$He targets
which provided measurements of the photo-absorption asymmetry $A_1$ for the
proton, deuteron and neutron respectively.
The measurements were performed by six experiments at CERN (EMC\cite{EMC}, SMC\cite{SMC}),
SLAC (E142\cite{142}, E143\cite{143}, E154\cite{154}) and DESY (HERMES\cite{HERMES}).
For each experiment we considered the data set 
with the most detailed $Q^2$ and $x$ binnings.
Therefore we did not include in the fit the $Q^2$-averaged data which are 
generally used for the determination of the polarised structure 
function $g_1(x)$. 
A total of 511 independent data points was included in the analysis: specifically 238 
proton data, 81  neutron data and 192 deuteron data.
The kinematic range covered by the data is 0.3 GeV$^2 < Q^2 <$ 70 GeV$^2$
and 4 GeV$^2 < W^2 <$ 300 GeV$^2$ for the proton and deuteron,
and 1 GeV$^2 < Q^2 <$ 15 GeV$^2$ and 4 GeV$^2 < W^2 <$ 70 GeV$^2$ for the
neutron.

The cross section difference $\Delta\sigma$ was derived from 
the measured virtual photo-absorption asymmetry $A_1$ and from the unpolarised 
structure function $F_1$:
\begin{equation}
\label{dsigma}
\Delta\sigma = 2A_1\sigma_T = \frac{8\pi^2\alpha}{m}\;\frac{A_1F_1}{K},
\end{equation}
 
\noindent where $\sigma_T$ is the total transverse cross section and
$K = \sqrt{\nu^2+Q^2}$ is the virtual photon flux factor.
The structure function $F_1 = F_2(1+\gamma^2)/(2x(1+R))$ was 
calculated from published parameterisations of the unpolarised structure 
functions $F_2$ \cite{NMCF2} for the proton, neutron and deuteron and 
of $R = \sigma_L/\sigma_T$, the ratio of the 
absorption cross sections \cite{WHITL} for longitudinally and transversely 
polarised virtual photons.  
The structure function R was assumed to be independent of the target
($R^p=R^n=R^d$).

The $\Delta\sigma$ experimental data were reproduced by a global 3-D fit
over the $Q^2$ and $W^2$ variables and the isospin dimension $T$:

\begin{equation}
\label{Dsigma}
\Delta\sigma = \Delta\sigma(T, W^2, Q^2)  =    
\left[2 a W^{2(\alpha^0_{a1}-1)} T + f W^{2(\alpha^0_{f1}-1)} + g \frac{lnW^2}{W^2}\right]
 ^{}F,
\end{equation}

\noindent where $F$ is the threshold factor

\begin{equation}
\label{threshold}
F = \left[\frac{W^2 - W_{\pi}^2}{W^2 - W_{\pi}^2 + Q^2 + m_r^2}\right]^p,
\end{equation}

\noindent where $W_{\pi}=1.12$ GeV is the single pion electro-production threshold and 
$m_r = 1.26$ GeV is a common mass scale of the exchanged particles which is
close to the $a_1$ and $f_1$ meson masses and to the two gluon exchange hadronic scale.
The exponent $p$ is given by

\begin{equation}
\label{power}
p = 1.5 \left(1 + \frac{Q^2}{Q^2 + M^2}\right),
\end{equation}

\noindent being $M$ a free parameter of the fit.
This threshold factor is similar to the ones used in the
description of unpolarised cross section $\sigma_T$ accounting for the behaviour of a 
cross section corresponding to a virtual particle with 
$W^2 > Q^2$[23-26].
$F$ increases from 0 at the pion electro-production threshold to $\approx 1$
at high $W^2$, when $W^2 \gg Q^2$.

Because Regge theory gives no indication about the relative weight and 
$Q^2$-dependence of the various contributions,
the three coefficients in eq.~\ref{Dsigma} were parametrised with
a simple and smooth $Q^2$-dependence, with two free parameters each :
 
\begin{equation}
\label{parameter}
a,f,g =p_1^{a,f,g} + p_2^{a,f,g}  t,
\end{equation}

\noindent where

\begin{equation}
\label{tdep}
t = log\frac{log \frac{Q^2 + Q_0^2}{\Lambda^2}}{log\frac{Q_0^2}{\Lambda^2}}.
\end{equation}
Here $\Lambda$ and $Q_0^2$ are the QCD cutoff and scale parameters, respectively.
We used $\Lambda = 0.255$ GeV and $Q_0^2 = 0.278$ GeV$^2$ in agreement with the ALLM 
parameterisation for the unpolarised total transverse cross section $\sigma_T$\cite{ALLM}.

The cross section difference for the proton and neutron were defined by eq. 5
with $T$=$+$1/2 and $T$=$-$1/2, respectively.
The cross section difference for the deuteron was defined $\Delta\sigma^{d}=
\left[\Delta\sigma^{p}+\Delta\sigma^{n}\right](1-1.5\omega_D)$ where 
$\omega_D=0.05$ is the D-wave probability in the deuteron\cite{mac}.


\section{Results}

In the following we describe three different models we used to reproduce the
$\Delta\sigma$ data. The results of the three models are presented in Table 1.

\noindent
{\bf Model I}

\noindent
We fitted proton, neutron and deuteron data using the parameterisation described in the
previous chapter and with
9 free parameters ($\alpha_{a1}^0, \alpha_{f1}^0, p_1^a, p_2^a, p_1^f, p_2^f, p_1^g, p_2^g, M^2$).
The fit reproduced the data with reduced $\chi^2/ndf =1.14$. 
The intercept of the $f_1$-meson trajectory was found within the expected range.
On the contrary, the intercept of the $a_1$-meson trajectory was not consistent with
the standard Regge theory.
This finding may imply that the $a_1$-meson trajectory is strongly not linear or not
parallel to the other meson trajectories (like the $a_2$-meson or the $\rho$-meson ones).

\noindent
{\bf Model II}

\noindent
To cure the problem of non-linearity or non-parallelism of the previous model
we introduced in the fit a smooth $Q^2$-dependence of $a_1$-meson intercept assuming it
equal to 0 for $Q^2$=0 and varying linearly with $t$.
With this assumption, the errors of the parameters of the coefficients 
and the $\chi^2/ndf$ were decreased.
The intercept of the $f_1$-meson trajectory was found in excellent agreement with the 
expectation.
The intercept of the $a_1$-meson trajectory changed of 0.25 in the $Q^2$ range 0-10
GeV$^2$. This behaviour is similar to the $Q^2$-evolution of the
reggeon and the pomeron intercepts found in the description of the unpolarised $\sigma_T$ data\cite{ALLM}.

\noindent
{\bf Model III}

\noindent
In this parameterisation, we considered an additional reggeon trajectory $2 r W^{2(\alpha^0_r-1)} T$
for the isovector part of the interaction, with no $Q^2$-dependence of the intercepts. 
The result of this Model are only slightly worse respect to the ones of Model II.
The intercept of the $f_1$-meson trajectory is within the 
expectation range. The intercept of the $a_1$-meson trajectory was found in excellent agreement 
with the expectation. The intercept $\alpha_r^0$ of the additional reggeon trajectory was found 
close to the $a_2$-meson intercept $\approx$ 0.5.

\begin{figure}[t]
\vspace{18cm}
\includegraphics{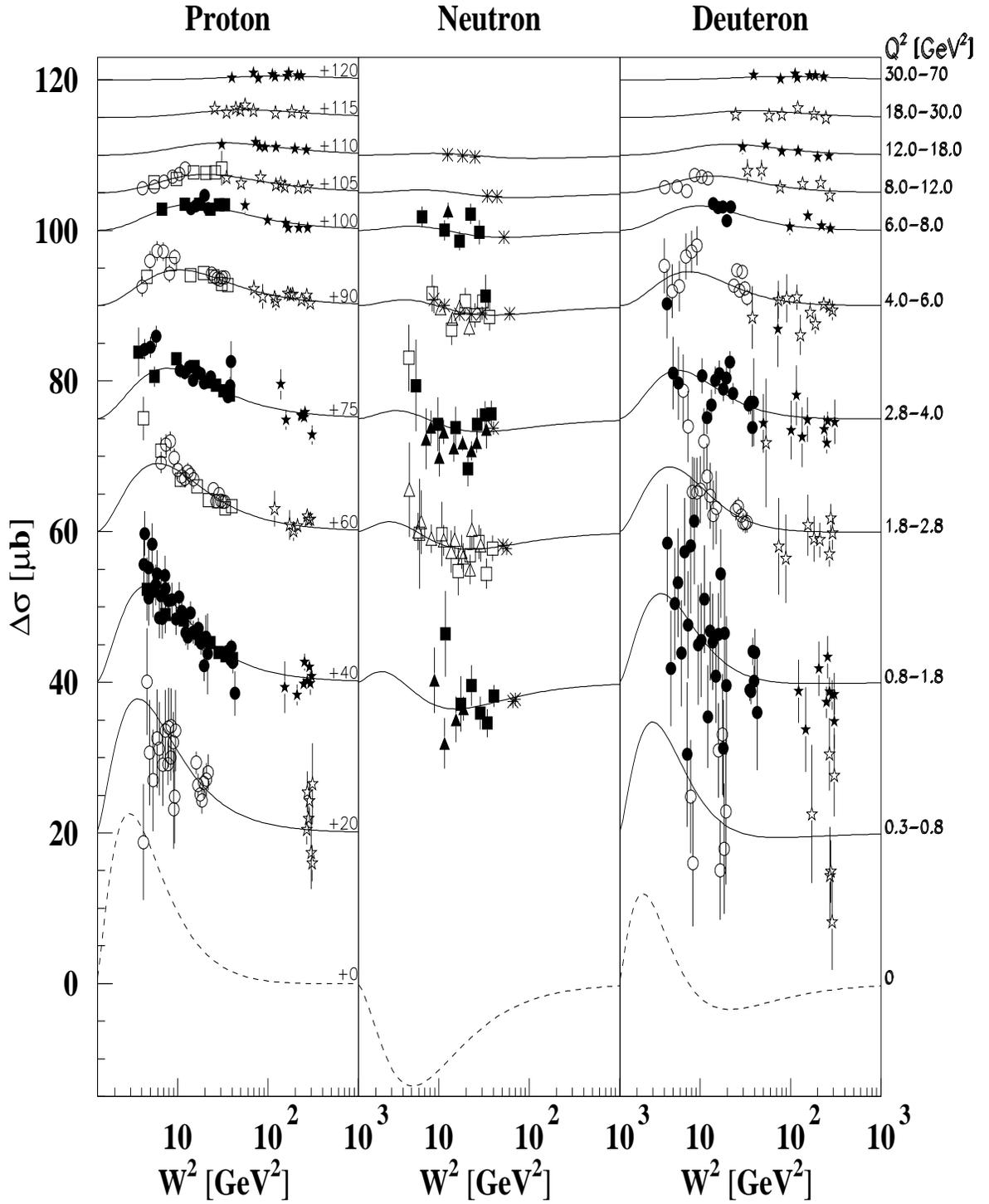}
\caption{Cross section differences as a function of $W^2$ for different $Q^2$-bins
(on the right scale).
Data are from Refs. 
\protect\cite{EMC,SMC} (stars),
\protect\cite{142} (triangles),
\protect\cite{143} (circles),
\protect\cite{154} (asterisks),
and \protect\cite{HERMES} (squares). 
For a better view, a $Q^2$-increasing constant value (on the right side of the
proton panel) has been added to the data
which are presented with open and close symbols alternatively.
Only data with statistical error lower than 10 $\mu$b are shown.  
The solid curves are the results of Model II calculated for the average $Q^2$-values
in each $Q^2$-bin.
The dashed curves are the predictions of Model II for $Q^2$=0.}
\label{fit}
\end{figure}

All the three Models provide a rather good description of the present data. The new and precise 
data for proton and deuteron that will be provided by the HERMES and E155\cite{155} collaborations will allow
to better discriminate between the above described approaches. 
In the following, we show the results for the Model II which provided the best $\chi^2/ndf$ and which
has the simplest phenomenological interpretation for the isovector component.
The comparison between the results of Model II and the experimental data as function of the three 
variables ($T$, $W^2$ and $Q^2$) is shown in fig.~\ref{fit}. 
For this purpose the data are presented grouped in $Q^2$-bins.

In fig.~\ref{fit} are also shown the results of Model II for $Q^2$=0.
The expectations for real photons are:
$\Delta\sigma^p$ is positive up to $\nu\sim 150$ GeV, ranging between 0 and $\sim23\mu$b;
$\Delta\sigma^n$ is negative and ranges between $\sim - 14$ and 0$\mu$b;
$\Delta\sigma^d$ ranges between $+$12 and $-4\mu$b being positive below $\nu\sim 3.5$ GeV.

\begin{figure}[t]
\vspace{10cm}
\includegraphics{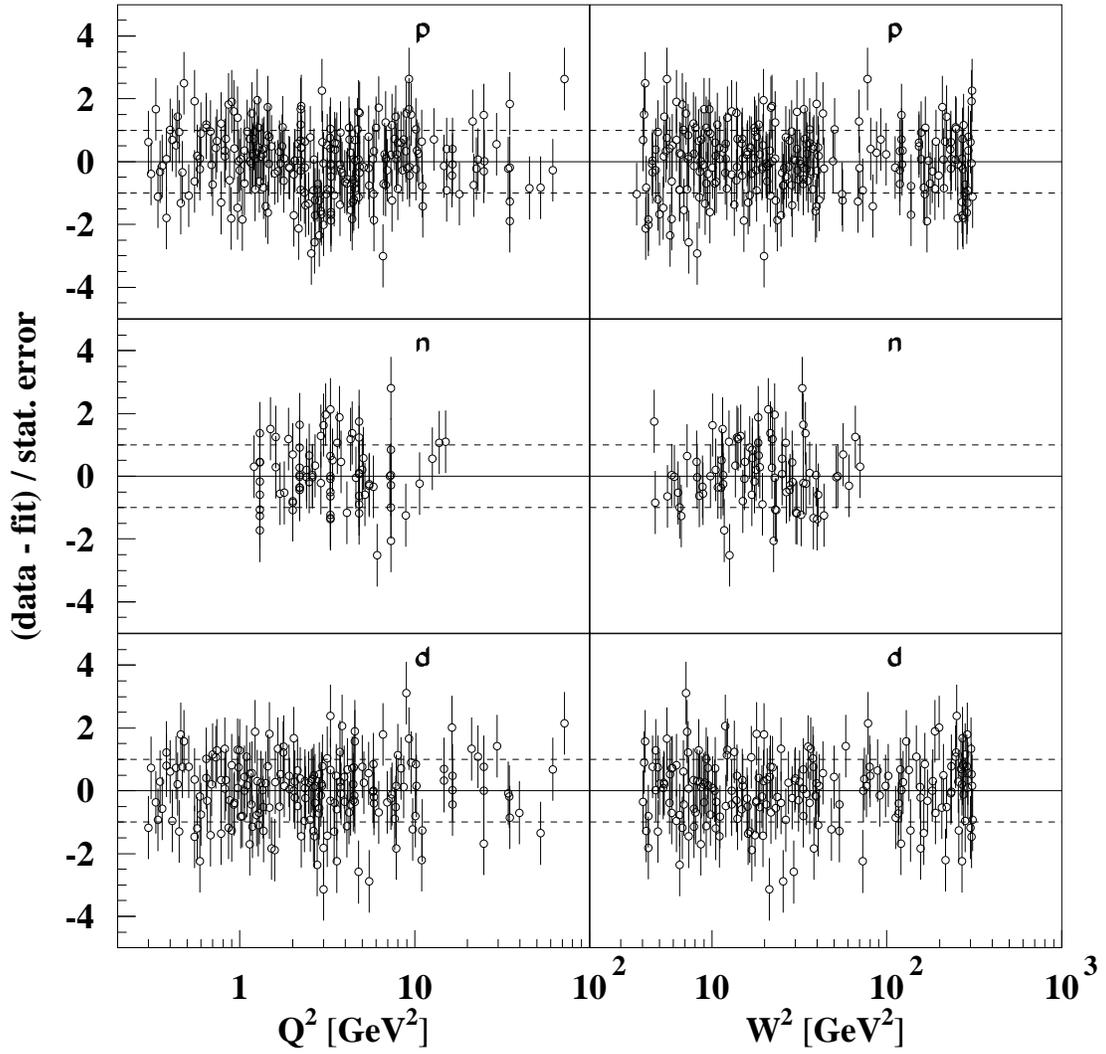}
\caption{Deviation of the experimental data from the curves of Model II, normalised by the
experimental statistical error, as function
of $Q^2$ and of $W^2$ for the proton, neutron and deuteron.}
\label{chi2}
\end{figure}

In fig.~\ref{chi2} the point to point difference between the fit of Model II and the experimental data is provided.
The difference has been normalised by the statistical error of each data point.
As it is seen the fit well reproduces all the data over the two orders of magnitude variation of
the $Q^2$ and $W^2$ variables.
The uniform distribution of this difference evidences the good quality of the fit 
and ensures a reliable extrapolation to $Q^2=0$.

\begin{figure}[t]
\vspace{12cm}
\includegraphics{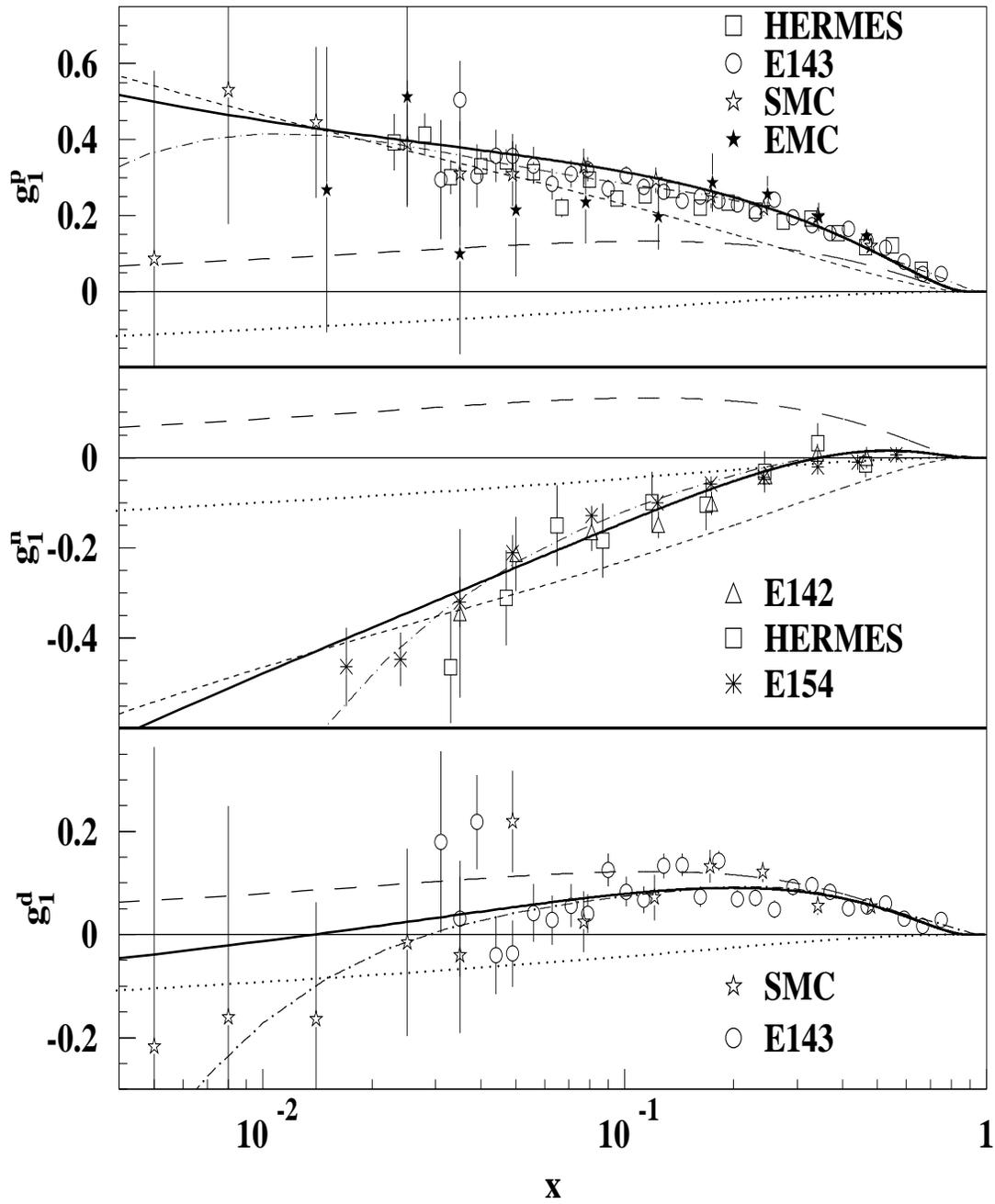}
\caption{Spin structure functions $g_1(x)$ at $Q^2=3$ GeV$^2$ (same notation as Fig. 1, for the data). 
The solid curves are the results of Model II. 
The short-dashed, long-dashed and dotted curves are the $a_1$, $f_1$ and two-gluons contributions
respectively.
The predictions of a NLO-QCD fit \protect\cite{GRSV} are indicated by the dot-dashed curves.}
\label{g1}
\end{figure}

A further comparison between Model II and experimental data is provided
in fig.~\ref{g1} where the expectations for the polarised structure functions
$g_1(x)$ are shown. These were evaluated at $Q^2 = 3$ GeV$^2$ using eqs.~\ref{dsigma} 
and ~\ref{Dsigma} for model II
under the approximation that $g_1(x,Q^2)\approx A_1(x)F_1(x,Q^2)$.
The curves are compared with $g_1(x)$ $Q^2$-averaged data evolved at the same $Q^2$ 
and with a NLO-QCD fit based on polarised parton density distributions\cite{GRSV}.
Our Model and the QCD model are in good agreement at least
for $x\geq$ 0.03 and both well reproduce the experimental data. 
At low-$x$, where they are different, the present experimental uncertainties are too large to provide
strong constraints for both models. 
Measurements of the spin structure functions at very low-$x$ using the HERA polarised
collider\cite{roe} will shed light on this subject. 

Also shown in fig.~\ref{g1} are the $a_1$-meson, $f_1$-meson and two-gluons exchange
individual contributions.
In this framework, the stronger experimental variation of $g_1^n$ at low-$x$ respect to $g_1^p$
is simply explained by the similar low-$x$ behaviors of the isovector and the isoscalar components
in the neutron.
The change of sign of $g_1^d$ is at $x \approx 0.015$ and is due to the 
different $x$-dependence of the two isoscalar components, being the $f_1$-meson
positive contribution dominating at large-$x$, while the two-gluons negative one at low-$x$.

\section{Contribution to the GDH integral}

\begin{figure}[t]
\vspace{8.5cm}
\includegraphics{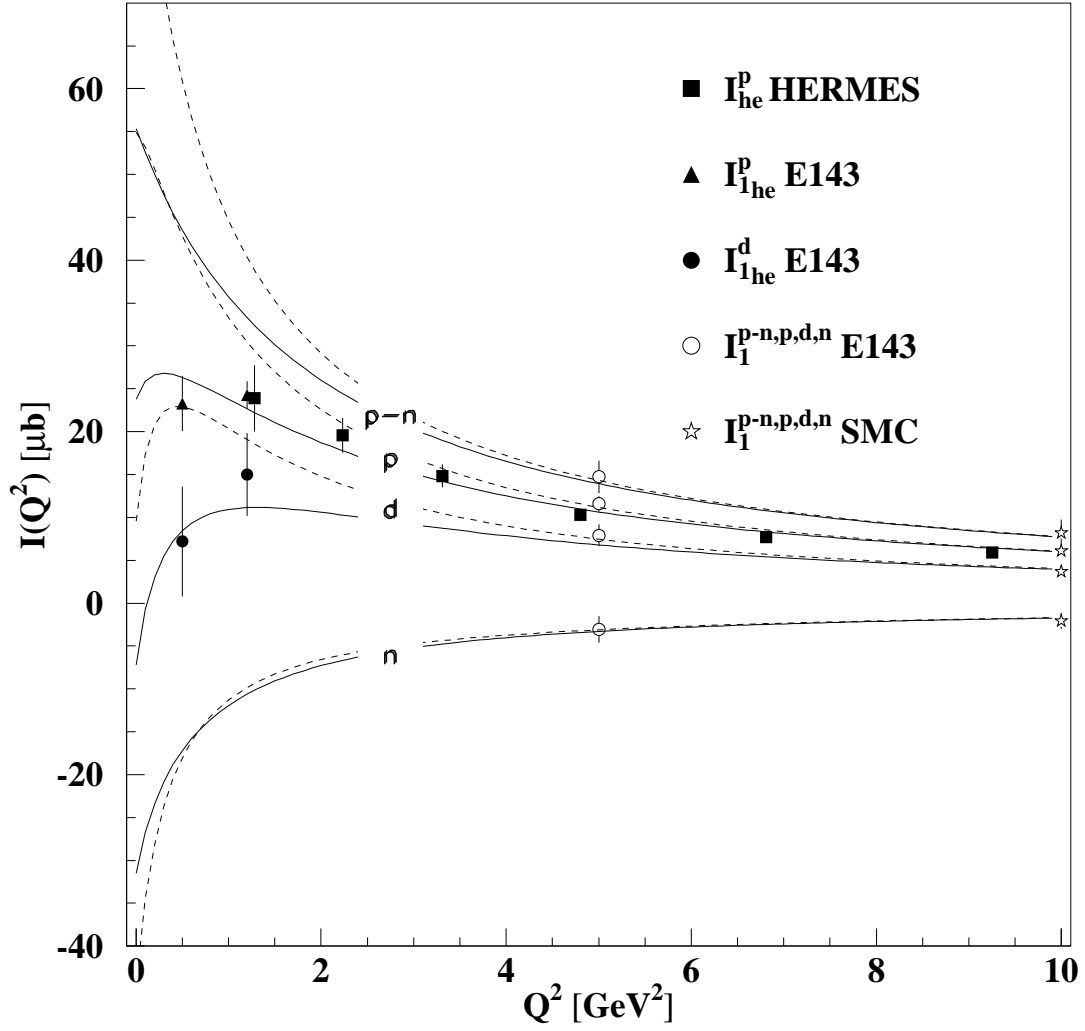}
\caption{$Q^2$-evolution of $I_{he}(Q^2)$ (solid curves) and of $I(Q^2)$ (dashed
curves) evaluated with Model II for (p-n), p, d and n.
The Model predictions are compared with experimental results.
The error bars show the quadratic combination of the statistical and the systematic uncertainties. }
\label{gdh}
\end{figure}

The $Q^2$-dependence of the high energy contribution ($W\ge2$ GeV) $I_{he}(Q^2)$
to the generalised GDH integral $I(Q^2)$ can be evaluated from eq.~\ref{gdhq2} with
a lower integration limit $\nu_0=(4-m^2+Q^2)/2m$ GeV.
In fig.~\ref{gdh} are shown the predictions for $I_{he}(Q^2)$ of Model II for the proton-neutron,
proton, deuteron and neutron. Our result for the proton well reproduce the recent HERMES measurements
\cite{HERMES}
performed at fixed values of $Q^2$ and in which their Regge extrapolation for $\nu > 23.5$ GeV
is taken into account. 
Also the E143 measurements\cite{143} of the high energy contribution $I_{1he}(Q^2)$ to the integral
$I_1(Q^2)$ for the proton and the deuteron are in reasonable agreement with our curves for
$I_{he}(Q^2)$.

In fig.~\ref{gdh} are also shown our results of the total integral $I(Q^2)=I_{he}(Q^2)+I_{le}(Q^2)$ 
which includes
the low energy contributions $I_{le}(Q^2)$ in the nucleon resonance region
($W<2$ GeV). As it is seen the high energy contribution 
is the dominant one at high $Q^2$.
At high $Q^2$, where the nucleon resonance excitation contribution is negligible and 
the approximation of eq.~\ref{gamma1} is valid, our results well agree with the
integrals $I_1$ measured by E143 at $Q^2=5$ GeV$^2$ \cite{143}
and by SMC at $Q^2=10$ GeV$^2$ \cite{SMC}.
  
The real photon limit $I_{he}(0)$ of the high-energy contribution to
the GDH integral for the three Models and 
for the proton, neutron and proton-neutron difference are listed in Table 1.
The three Model predictions are consistent among each other within $\sim 10\%$.
Our predictions for the proton and deuteron are in agreement with the predictions $I_{he}^p =$ 25$\pm$10
$\mu$b and $I_{he}^d \sim 0$ $\mu$b from ref.\cite{bass} in which a Regge-inspired approach was also used.
In Table 1 the evaluations of the low-energy parts $I_{le}(0)$ are also presented. The
latter results are
more model dependent respect to the high-energy contributions and are strongly affected by
the choice of the threshold factor $F$. Moreover $I_{le}(0)$ is
less constrained by the experimental data and therefore provides only an approximate estimate
of the multi-hadron and non-resonant photo-production contribution at low energy.

In Table 2 the $I_{he}(0)$ and $I_{le}(0)$ contributions, averaged over the three models, are
reported. 
The quoted errors for the integrals were derived considering both the 
contribution of the free parameter errors and the contribution of the
fit parameterisation. The latter, which is the dominant one for the real
photon extrapolation, has been
evaluated from the slightly different predictions of
the three models,
from varying the fixed parameters of the fit ($Q_0^2$, $\Lambda^2$, $m_r^2$) 
within a factor two around the chosen values and 
from a $Q^2$-independent choice of the power expression of the threshold function 
$F$.

In Table 2 we also reported the contributions from the most recent analysis of single-pion 
photo-production in both resonant and non-resonant channels \cite{dr} and 
from the decay of the
nucleon resonances in two-pions \cite{ka}. As it is seen the discrepancies between the sum of these
two latter contributions and the GDH sum rule expectation is reduced by a factor $\geq$ 2 if we
add our results for the high-energy contributions.
Moreover the above discrepancies
are almost canceled if we also include our estimates for the low-energy region. 

Therefore the evaluation of multi-hadron photoproduction processes accounts for 
a relevant fraction of the GDH expectations for the proton and the neutron. In addition it
represents the largest contribution for the isovector sum rule, in which the resonance
contribution is mostly canceled.
These phenomenological predictions will be soon tested by
ongoing and upcoming experiments with real \cite{bia} and virtual photons of low $Q^2$ \cite{EX97}.

\section{Conclusions}

We have studied the deep inelastic scattering of longitudinally polarised leptons off
polarised targets in terms of polarised cross sections.
The cross sections have been fitted by three different parameterisations in which the isovector
and the isoscalar contributions from Regge inspired models have been considered.   

With these parameterisations we were able to well describe, over the whole kinematic
range of $Q^2$ and $W^2$, the experimental polarised cross sections $\sigma_{1/2}-\sigma_{3/2}$
and structure functions $g_1$ for the proton, neutron and deuteron and the relevant
integrals.
We were able to evaluate the $Q^2$-evolution of the high energy contribution ($W\ge$ 2 GeV) 
to the generalised GDH integral and to provide an accurate estimate for this contribution in the real photon
limit. The latter contribution was found to be about one half of the present discrepancy between
previous multipole analysis and the GDH sum rule expectations.
Moreover we also provided a rough estimate of the low energy ($W<$ 2 GeV) multi-hadron and non-resonance contribution
to the GDH integral.
 
At present, while waiting for direct and experimental verifications, from the informations
coming from unpolarised single-pion photo-production data at low energy and from inclusive polarised
electro-production data at high energy,  
it seems reasonable to conclude that no room is left for a large violation of the GDH sum rule.

\section{Acknowledgements}
We would like to thank S.D.~Bass for many suggestions and for reading the paper and G.~Pancheri and K.~Oganesyan 
for useful discussions.
This work was partially supported by the TMR contribution (contract FMRX-CT96-0008) from the European Community.

\newpage

\begin{table}
\label{table1}
\begin{center}
\caption{Parameter values and contributions to the GDH integral provided by
the fit of the data with the different models. The error estimates have been
obtained from a MINOS analysis of the MINUIT program.}
\begin{tabular}{lc@{\hspace{0.2cm}}c@{\hspace{0.2cm}}c}
\hline\hline
                                &Model I        &Model II               &Model III \\
\hline
$p_1^a$ [$\mu$b]                &110$\pm$5      &135$\pm$5              &32 $\pm$34     \\
$p_2^a$ [$\mu$b]                &-41$\pm$4      &-75$\pm$3              &135 $\pm$48    \\
$p_1^f$ [$\mu$b]                &80$\pm$28      &110$\pm$23             &120$\pm$14     \\
$p_2^f$ [$\mu$b]                &165$\pm$47     &84$\pm$31              &62$\pm$14      \\
$p_1^g$ [$\mu$b]                &-21$\pm$2      &-26$\pm$3              &-32$\pm$2      \\
$p_2^g$ [$\mu$b]                &20$\pm$2       &23$\pm$2               &24$\pm$2       \\
$p_1^r$ [$\mu$b]                &-              &-                      &41$\pm$2       \\
$p_2^r$ [$\mu$b]                &-              &-                      &-20$\pm$1      \\
$M^2$   [GeV$^2$]               &3.7$\pm$0.7    &4.8$\pm$0.8            &2.9$\pm$0.5    \\
$\alpha_{a_1}^0$                &0.16$\pm$0.02  &$(0.20\pm0.02)\cdot t$ &-0.27$\pm$0.03 \\
$\alpha_{f_1}^0$                &-0.49$\pm$0.10 &-0.37$\pm$0.09         &-0.51$\pm$0.15 \\
$\alpha_r^0$                    &-              &-                      &0.43$\pm$0.02  \\
\hline
$\chi^2/ndf$                    &1.14           &1.11                   &1.12   \\
\hline
$I^p_{he}(0)$ [$\mu$b]          &$+28$          &$+24$                  &$+27$          \\
$I^n_{he}(0)$ [$\mu$b]          &$-40$          &$-32$                  &$-34$          \\
$I^d_{he}(0)$ [$\mu$b]          &$-10$          &$-7$                   &$-6$           \\
$I^{p-n}_{he}(0)$ [$\mu$b]      &$+68$          &$+55$                  &$+61$          \\
$I^p_{le}(0)$   [$\mu$b]        &$+27$          &$+32$                  &$+25$          \\
$I^n_{le}(0)$   [$\mu$b]        &$-16$          &$-14$                  &$-3$           \\
$I^d_{le}(0)$   [$\mu$b]        &$+10$          &$+17$                  &$+20$          \\
$I^{p-n}_{le}(0)$ [$\mu$b]      &$+43$          &$+45$                  &$+28$          \\
\hline\hline
\end{tabular}
\end{center}

\label{table2}
\begin{center}
\caption{Evaluation of the GDH integral $I_{TOT}$ for the proton, neutron and
proton-neutron difference and comparison with the relevant GDH sum rule expectations.
$I_{TOT}$ is computed by the sum of the predicted contributions for single-pion
photo-production\protect\cite{dr}, double-pion photo-production from the decay of the nucleon resonances 
$N^{*}$\cite{ka}, and for the high-energy ($W\geq2$ GeV) and low-energy ($W<2$ GeV) multi-hadron
photo-production (this work).}
\begin{tabular}{lccc}
\hline\hline
                                        &$I_p$ [$\mu$b] &$I_n$ [$\mu$b] &$I_{p-n}$ [$\mu$b]     \\
\hline
$I_{N \pi}$\cite{dr}                    &$-196$         &$-145$         &$-51$  \\
$I_{N^{*}\rightarrow N\pi\pi}$\cite{ka} &$-65$  &$-35$          &$-30$  \\
$I_{he}(0)$ this work                   &$+26\pm7$      &$-35\pm11$     &$+61\pm12$     \\
$I_{le}(0)$ this work                   &$+28\pm19$     &$-11\pm14$     &$+39\pm29$     \\
\hline
$I_{TOT}$                               &$-207\pm23$    &$-226\pm22$    &$+19\pm37$     \\
\hline
$GDH$ sum rule                          &$-204$         &$-233$         &$+29$  \\
\hline\hline
\end{tabular}

\end{center}
\end{table}

\vfill
\eject

\vfill
\eject

\vfill
\eject

\vfill
\eject

\end{document}